\documentclass[a4paper,fleqn,usenatbib]{mnras}

\usepackage[T1]{fontenc}
\usepackage{ae,aecompl}

\usepackage{graphicx}	
\usepackage{amsmath}	
\usepackage{amssymb}	

\def\msun{\hbox{M$_\odot$}}

\title[eMSTO star clusters in the LMC]{An Analysis of the Population of Extended Main Sequence Turn-off Clusters in the Large Magellanic Cloud}

\author[Piatti \& Bastian]{
Andr\'es E. Piatti$^{1,2}$\thanks{E-mail: andres@oac.unc.edu.ar (AEP)}
and Nate Bastian$^{3}$
\\
$^{1}$Observatorio Astron\'omico, Universidad Nacional de C\'ordoba, Laprida 854, 5000, 
C\'ordoba, Argentina\\
$^{2}$Consejo Nacional de Investigaciones Cient\'{\i}ficas y T\'ecnicas, Av. Rivadavia 1917, 
C1033AAJ, Buenos Aires, Argentina\\
$^{3}$Astrophysics Research Institute, Liverpool John Moores University, 146 Brownlow Hill, Liverpool L3 5RF, UK
}

\date{Accepted XXX. Received YYY; in original form ZZZ}

\pubyear{2015}

\begin{document}
\label{firstpage}
\pagerange{\pageref{firstpage}--\pageref{lastpage}}
\maketitle

\begin{abstract}
We combine a number of recent studies of the extended main sequence turn-off (eMSTO) phenomenon in intermediate age stellar ($1-2$~Gyr) clusters in the Large Magellanic Cloud (LMC) in order to investigate its origin.  By employing the largest sample of eMSTO LMC clusters so far used,
we show that cluster core radii, masses, and dynamical state are not related to the
genesis of eMSTOs. Indeed, clusters in our sample have core radii, masses and age-relaxation
time ratios in the range $\approx$ 2--6 pc, 3.35-- 5.50 (log($M_{cls}$/$M_\odot$) and 0.2--8.0, respectively. These results imply that the eMSTO phenomenon is not caused by actual age spreads 
within the clusters. Furthermore, we confirm from a larger cluster sample recent results
including young eMSTO LMC clusters, that the FWHM at the MSTOs correlates most strongly with 
cluster age, suggesting that a stellar evolutionary effect is the underlying cause. 
\end{abstract}




\section{Introduction}

It is now clear that a large fraction of studied massive stellar clusters in the Large Magellanic Cloud (LMC) host stellar populations which display a spread in the main sequence turn-off (MSTO) that cannot be explained by photometric errors or stellar binarity 
\citep[e.g.][]{mietal09}.  From these extended MSTOs (eMSTOs), one can infer age spreads within the clusters, of the order of 100-700~Myr \citep[e.g.][]{maetal08,rubeleetal2011,rubeleetal2013}. 
If such age spreads are real, this would dramatically affect our understanding of the formation of stellar clusters, which are generally thought to be classic simple stellar populations (SSPs).  In addition to actual age spreads, stellar evolutionary effects (e.g., due to stellar rotation) have been invoked to explain the eMSTOs,  which may mimic the effect of an age spread 
\citep[e.g.][]{bdm09,bastianetal16}.  Hence, the cause of the eMSTO has received particular attention and support
by different research groups.

A number of works have suggested correlations between the inferred age spreads and physical properties of the clusters. The idea behind these suggestions is
that some property of the cluster (e.g., mass, radius, density or escape velocity) 
either determines whether it can undergo
extended star formation, or is otherwise related to such an event. For example, \citet{kelleretal11} and \citet{kelleretal12} 
have suggested that  a large core radius is
a by-product of the process by which an eMSTO is formed. In addition, 
\citet{goudfrooijetal11,goudfrooijetal11b} 
and \citet{goudfrooijetal14} have suggested that cluster mass is the property responsible, with only massive clusters (i.e., those with high escape velocities) being able to form multiple generations of stars. 

Alternatively, other studies have suggested that true age spreads do not exist within the clusters, with the eMSTOs being caused by stellar evolutionary effects \citep[e.g.][]{bastianetal16}.  In this case, we would not expect correlations with a physical property of the clusters, but instead would expect a correlation between cluster age and the inferred age spread \citep{bh15}. In fact, such a correlation exists which is consistent with the predictions of rotating stellar isochrones 
\citep{niederhoferetal15a,niederhoferetal15b}.  
The observed correlation between the inferred age spread and the age of the cluster is the strongest evidence so far against the true age spread scenario.  Additionally, a number of studies have searched for multiple star-forming events in younger massive clusters, in which such age spreads should be 
extremely clear, and to date, no evidence for such extended star-formation histories
 have been found 
\citep[e.g.][]{bastianetal13,cabrera-zirietal14},\citep{cabrera-zirietal16,niederhoferetal16}.

In this paper we combine the catalogues of \citet{goudfrooijetal14} of massive clusters in the LMC hosting eMSTOs with that from our own studies 
\citep{p13,petal14,pb16}, 
in order to 
bring more evidence about that the eMSTO is linked to stellar evolutionary origins. 
In \S~\ref{sec:obs} we introduce the data sets and used them to derive cluster structural
parameters and masses to explore whether cluster properties are responsible for the eMSTO phenomenon. We discuss the results of our analysis in \S~\ref{sec:discussion}.  Our conclusions 
are presented in \S~\ref{sec:conclusions}.

\section{Cluster astrophysical properties}
\label{sec:obs}
\subsection{Structural parameters}

The cluster sample analysed here consists in 14 LMC clusters observed by \citet{petal14}, 2 LMC 
clusters studied by \citet{pb16} and SL\,529 \citep{p13}. Sixteen clusters were imaged with the 
Gemini telescope through the GMOS-S instrument and the $g,i$ filters attached. SL\,529 was observed through the $g,r$ filters instead. These set of observations represent an homogeneous sample 
not only  from an observational point of view, but also because the cluster ages and related fundamental parameters were derived using the same techniques and procedures.

\citet{petal14} presented their photometric data sets with the aim of investigating the frequency of occurrence
of the eMSTO phenomenon in the LMC cluster population through the link of eMSTOs to large core radii \citep{kelleretal12}. They selected
an age- and luminosity-limited sample with the original aim of examining if
the formation of multiple populations is a general phase of cluster
evolution, one possibly related to the puzzling multiple populations
seen in ancient Galactic globular clusters \citep[see, for example][]{carrettaetal10}.
The detailed analysis on the presence or absence of eMSTOs and their connexion to large
core radii in these clusters is conducted in this work. On the other hand, the 2 LMC 
clusters studied by \citet{pb16} and SL\,529 have confirmed eMSTOs, and were simply added to the
analysed cluster sample.

We derive cluster structural parameters for those clusters in \citet{petal14} and SL\,529
from a fully consistent method to that described 
by \citet{pb16} for the low-mass LMC clusters LW\,477 and LW\,483. Briefly, we determined the geometrical 
centres of the clusters by fitting Gaussian distributions to the star counts in the $x$ and $y$ 
directions, as well as directly on their deepest $g,i$
images, and recovered  independent cluster central coordinates closer than $\pm$10 pixels 
($\sim$ 1.5 arcsec). We used the resulting centres to checked whether the clusters exhibit some 
elliptical signature and derived ellipticity values in the range 0.00--0.20 $\pm$ 0.05 with an 
average of 0.08 $\pm$ 0.06.

\begin{figure*}
        \includegraphics[width=\columnwidth]{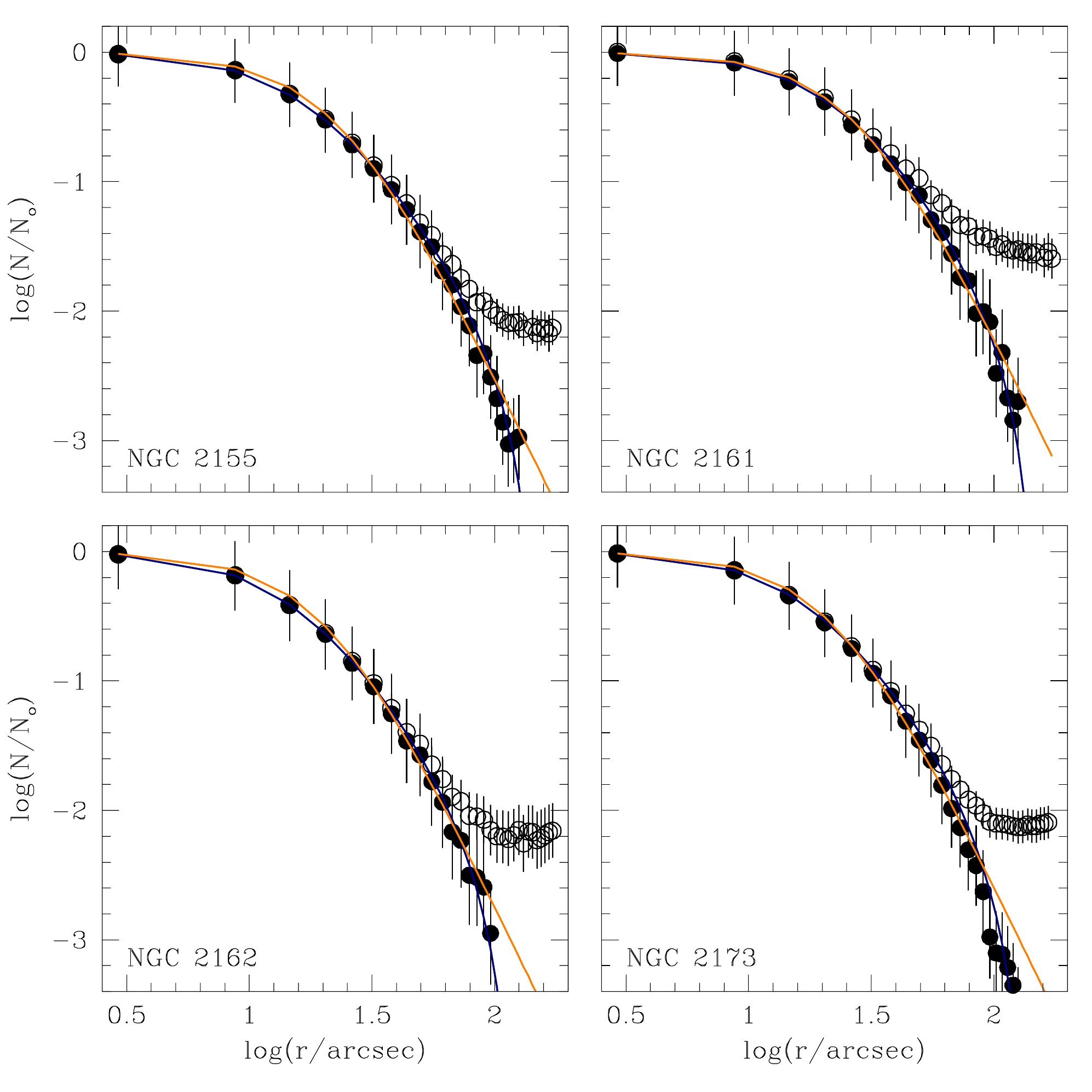}
        \includegraphics[width=\columnwidth]{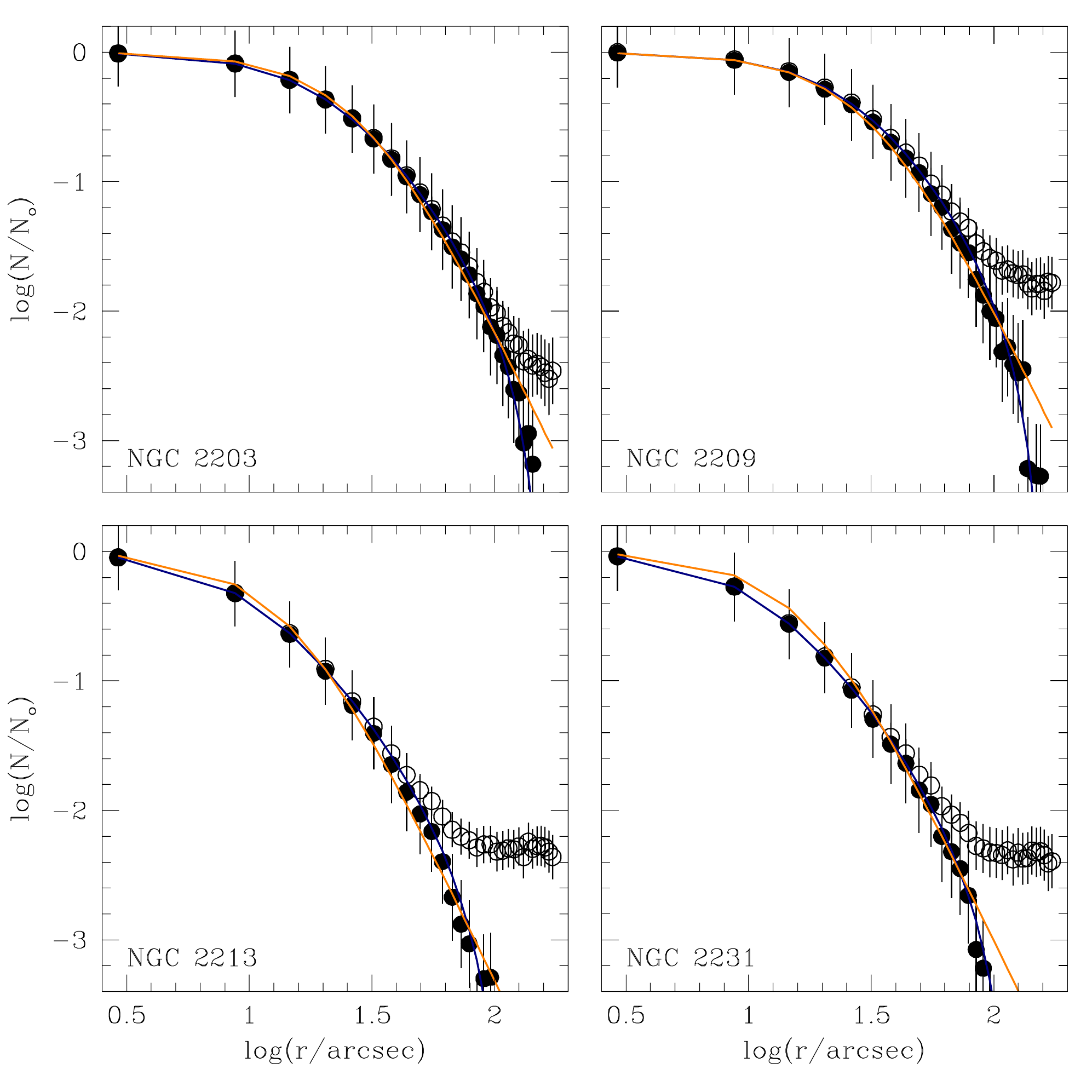}
        \includegraphics[width=\columnwidth]{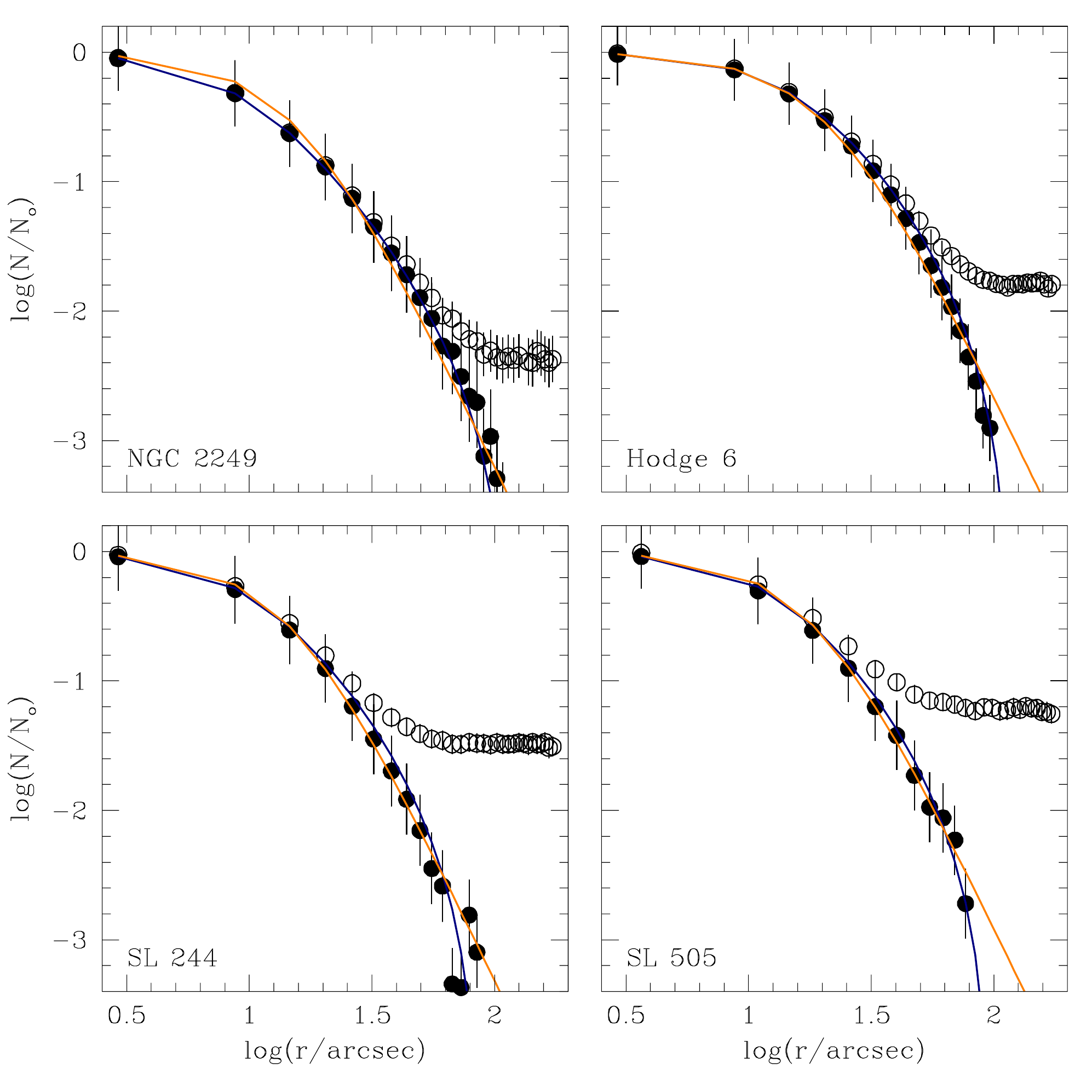}
        \includegraphics[width=\columnwidth]{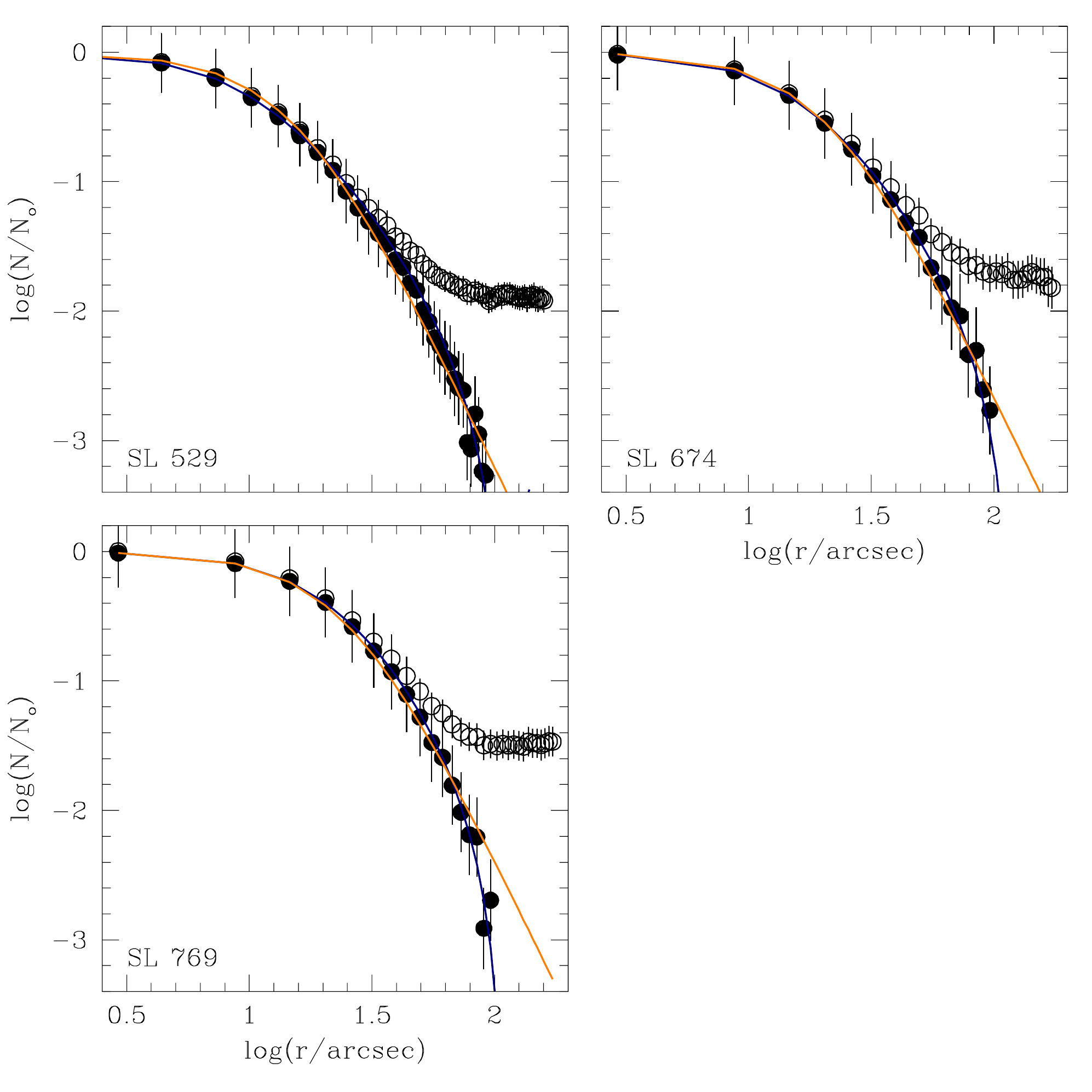}
    \caption{Stellar density cluster profiles obtained from star counts for the cluster sample. Open and filled circles refer to measured and
background subtracted surface brightness profiles, respectively. Blue and orange solid lines
depict the fitted King and Plummer curves, respectively.}
    \label{fig:fig1}
\end{figure*}

\begin{figure}
        \includegraphics[width=\columnwidth]{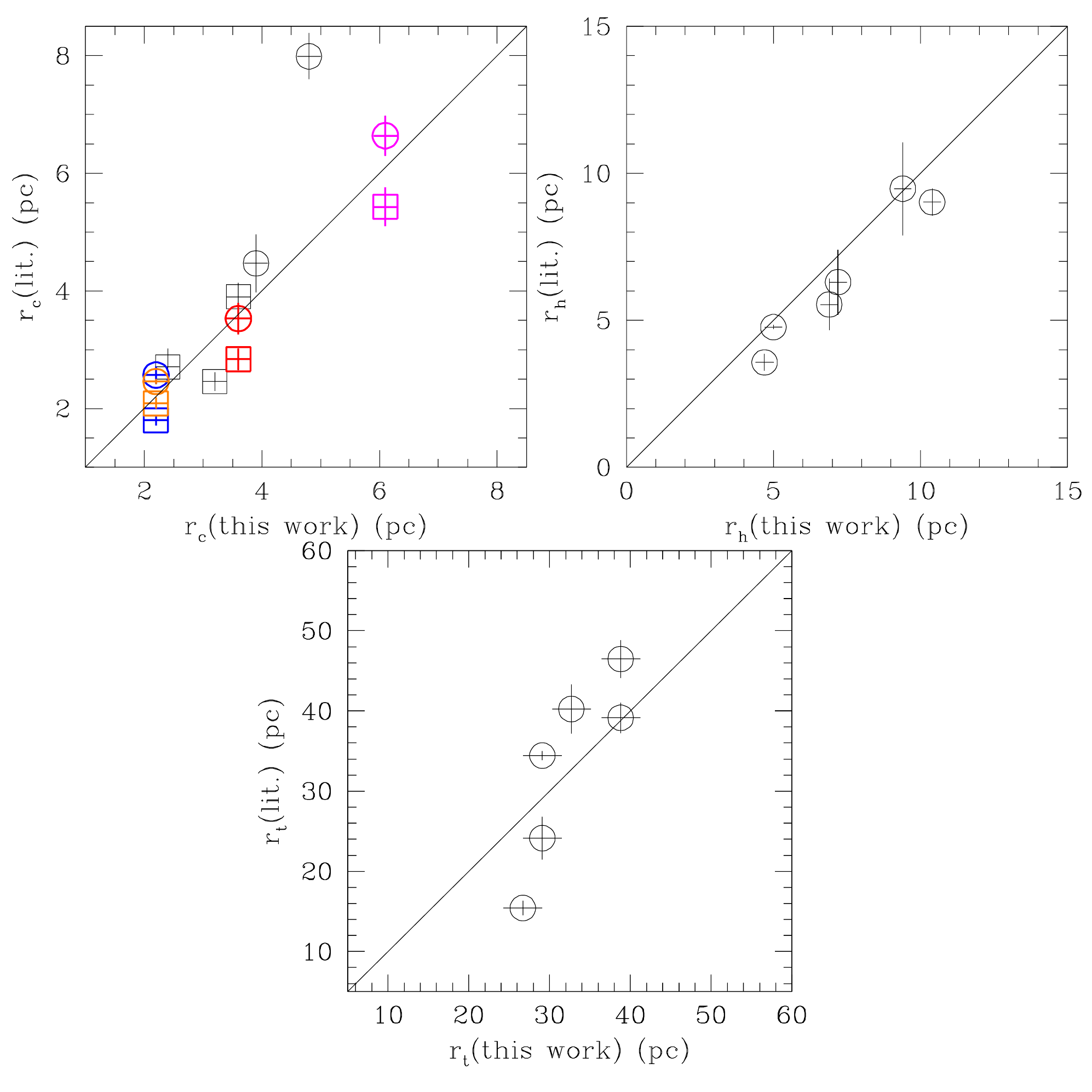}
    \caption{Comparison of structural parameters derived in this work with
those published by \citet{cetal14} and \citet{goudfrooijetal14} are
drawn with open circles. Errorbars for the individual values are also drawn. The solid lines represent the identity relation. Core radii derived by \citet{mg03} are represented by
open boxes. Coloured symbols refer to NGC\,2173 (red), 2209 (magenta), 2213 (blue) and 2249 (orange).}
    \label{fig:fig2}
\end{figure}

For the clusters in our sample, we built 
the cluster stellar density profiles --expressed as number of stars per arcsec$^2$-- from
the completeness corrected star counts performed using the cluster photometric catalogues by \citet{p13} and \citet{petal14}. The background corrected stellar density profiles were 
fitted using both \citet{king62} and  \citet{plummer11} models; the later is used to derived an independent
estimate of each clusters half-mass radius.  The resulting density profiles are shown in 
Fig.\ref{fig:fig1}. In the figure, we represented the constructed and background subtracted stellar density profiles with
open and filled circles, respectively. Errorbars represent Poisson
errors, to which we added the rms error of the background star
count to the background subtracted density profiles.
The derived core ($r_c$), half-mass 
($r_h$) and tidal ($r_t$) radii are listed in Table~\ref{tab:table1}. 
As is seen in Fig.~\ref{fig:fig2}, there is an overall very good agreement
of our core, half-mass and tidal radii with those derived by \citet{cetal14} and 
\citet{goudfrooijetal14}  (open circles) for a total of 6 clusters in common (NGC\,2173, 2203, 
2209, 2213, 2249 and Hodge\,6).  NGC\,2203 is the most discrepant point
in the $r_c$ panel, although its $r_h$ value agrees well with previous determinations. The $r_t$ panel shows
relatively more scatter, which could be due to the fact that the HST field is not wide
enough as to trace accurately the outer cluster regions. We also compared our $r_c$ values
with those of \citet{mg03} (open boxes) for 7 clusters in common, namely: NGC\,2155, 2162, 2173
(red), 2209 (magenta), 2213 (blue), 2231 and 2249 (orange).

\begin{figure}
\includegraphics[width=\columnwidth]{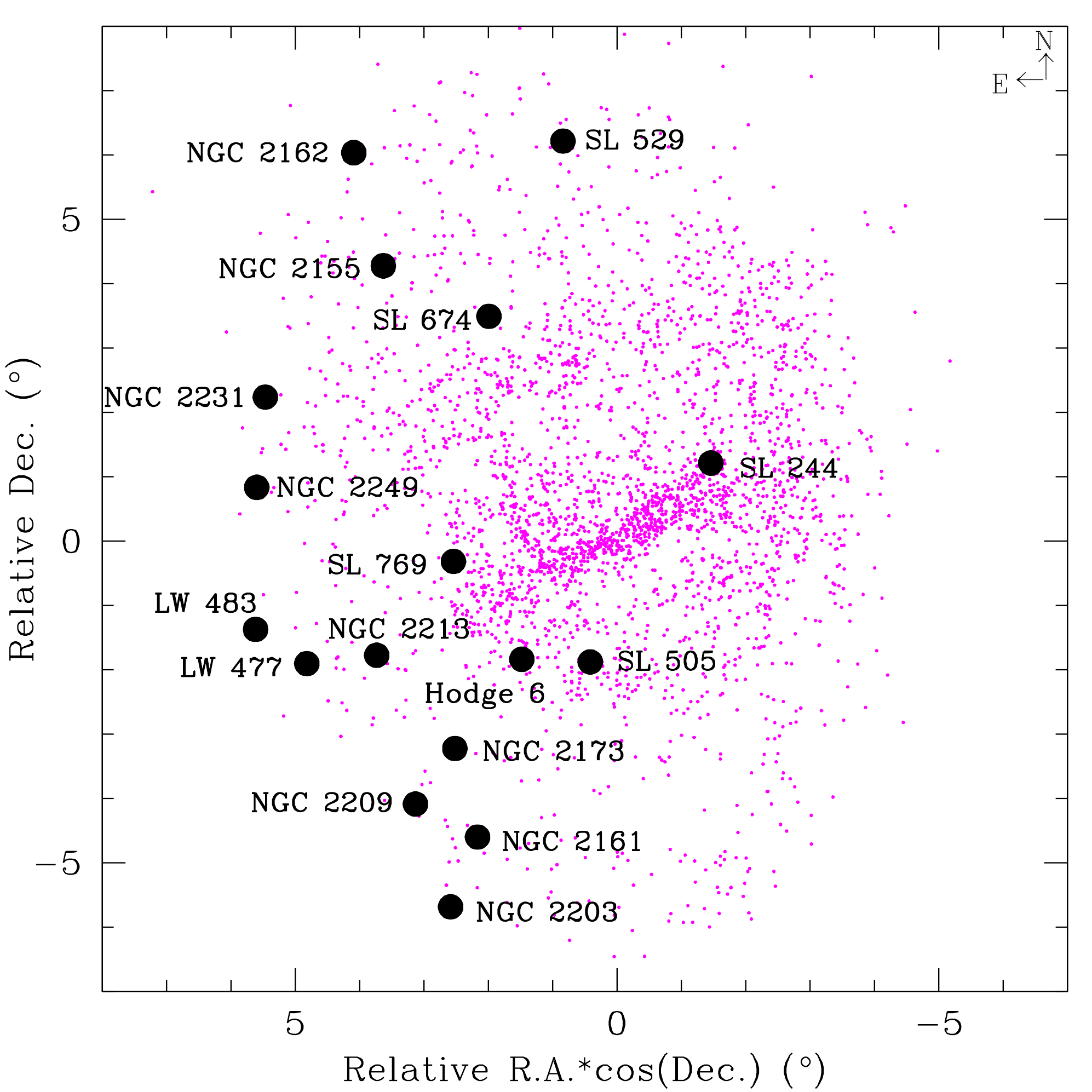}
\caption{Spatial distribution of the Bica et al. (2008)'s catalogue of star
clusters in the LMC centred at R.A. = 05$^h$ 23$^m$ 34$^s$ , Dec. = -69$\degr$ 
45$\arcmin$ 22$\arcsec$ (J2000), projected onto the sky, in magenta-coloured filled circles.
The studied clusters are highlighted
with black filled circles.}
\label{fig:fig3}
\end{figure}

The masses of the clusters in our sample were derived by comparing the observed integrated magnitude 
of each cluster (corrected for distance modulus and extinction) to that of the \citet{metal08} SSP 
models with a \citet{kroupa02} stellar initial mass function and a metallicity of $Z=0.008$ of the 
appropriate age.  The errors in the mass estimate are driven largely by the choice of models (which were 
also used in estimating the ages).  We estimate the uncertainty in the mass to be 
$\sigma$$(\log(M_{cls}/M_{\odot}))$ $\sim0.3$~dex, i.e., about a factor of two.
From the derived masses we estimated both the Jacobi tidal radius \citep[$r_J$,][]{cw90} and the 
half-mass relaxation time \citep[$t_r$,][]{sh71} of each cluster. We refer the reader to \citet{pb16}
for details on the values of the various parameters involved. Table~\ref{tab:table1} lists the
resulting $M_{cls}$,  $r_J$ and $t_r$ values. The latter are in excellent agreement with those coming from
using Fig. 20 of \citet{pijlooetal15}. Indeed, we found no difference between them, independently 
if \citet{pijlooetal15}'s ones were interpolated by considering 
scenarios with and without mass segregation. Additionally, we found a half-mass density range of 
3-95 $M_{\odot}$ pc$^3$ for the cluster sample. These value are much larger than the minimum density 
a cluster needs to have in order to be stable against the tidal disruption of a galaxy 
($\sim$ 0.1$M_{\odot}$ pc$^{3}$, \citet{bok34}). Accordingly, \citet{wilkinsonetal03} also showed that 
the tidal field of the LMC does not cause any perturbation on the clusters.
For the sake of the readers Fig.~\ref{fig:fig3} shows the spatial distribution of the cluster
sample.
 By comparing the resulting
$r_t$ and $r_J$ values, we found however that SL\,244, 505, 674, 769 and
LW\,477 have tidal radii larger than the respective Jacobi's ones \citep[see][]{getal11}.

\begin{table*}
\caption{Fundamental properties of LMC star clusters.}
\label{tab:table1}
\begin{tabular}{@{}lcccccccccc}\hline
Star cluster &$\log(t)$ &  $r_c$ & $r_h$ & $r_t$ & $r_J$ & $M_{cls}$    & $t_r$ & $FWHM$$_{\rm MSTO}$ & eMSTO$_{\rm lit}$ & Ref.\\
             & & (pc)  & (pc)  & (pc)  & (pc)  & (10$^3$~$M_{\odot}$) & (Myr) & (Myr) & (Myr) & \\\hline
NGC\,2155 &9.50$\pm$0.05 &3.6$\pm$0.2 & 7.6$\pm$0.3 & 36.4$\pm$2.4 & 48.6 & 53.0$\pm$36.6 & 985 & 130$\pm$127 & &\\
NGC\,2161 &9.35$\pm$0.05 &4.8$\pm$0.2 & 9.1$\pm$0.3 & 36.4$\pm$2.4 & 42.2 & 22.0$\pm$15.2 & 925 & 208$\pm$45 & &\\
NGC\,2162 &9.20$\pm$0.05 &3.2$\pm$0.2 & 6.6$\pm$0.3 & 29.1$\pm$2.4 & 55.1 & 37.5$\pm$25.9 & 702 & 225$\pm$25 & &\\
NGC\,2173 &9.25$\pm$0.05 &3.6$\pm$0.2 & 7.2$\pm$0.3 & 32.7$\pm$2.4 & 51.3 & 63.9$\pm$44.2 & 996 & 342$\pm$49 & 431 & 2\\
NGC\,2203 &9.30$\pm$0.05 &4.8$\pm$0.2 & 9.4$\pm$0.3 & 38.8$\pm$2.4 & 87.9 &111.1$\pm$76.7 &1855 & 349$\pm$15 & 475 & 2\\
NGC\,2209 &9.15$\pm$0.05 &6.1$\pm$0.2 &10.4$\pm$0.3 & 38.8$\pm$2.4 & 47.5 & 26.2$\pm$18.1 &1202 & 371$\pm$41 & 230 & 1\\
NGC\,2213 &9.25$\pm$0.05 &2.2$\pm$0.2 & 4.7$\pm$0.3 & 26.7$\pm$2.4 & 46.6 & 42.4$\pm$29.3 & 445 & 356$\pm$26 & 329 & 2\\
NGC\,2231 &9.20$\pm$0.05 &2.4$\pm$0.2 & 5.7$\pm$0.3 & 29.1$\pm$2.4 & 45.9 & 25.7$\pm$17.8 & 481 & 300$\pm$17 & &\\
NGC\,2249 &9.15$\pm$0.05 &2.2$\pm$0.2 & 5.0$\pm$0.3 & 29.1$\pm$2.4 & 44.6 & 21.0$\pm$14.5 & 372 & 174$\pm$20 & 450 & 1\\
Hodge\,6  &9.40$\pm$0.10 &3.9$\pm$0.2 & 6.9$\pm$0.3 & 29.1$\pm$2.4 & 33.4 & 79.9$\pm$55.2 &1019 & 235$\pm$75 & 238 & 2\\ 
SL\,244   &9.40$\pm$0.10 &2.4$\pm$0.2 & 4.7$\pm$0.3 & 21.8$\pm$2.4 & 11.2 & 24.5$\pm$16.9 & 358 & 285$\pm$27 & &\\
SL\,505   &9.30$\pm$0.10 &3.2$\pm$0.2 & 6.0$\pm$0.3 & 24.2$\pm$2.4 & 13.1 & 10.9$\pm$7.5  & 374 & 458$\pm$41 & &\\
SL\,529   &9.35$\pm$0.05 &2.4$\pm$0.2 & 5.0$\pm$0.3 & 26.7$\pm$2.4 & 31.1 & 10.0$\pm$6.9  & 280 & 406$\pm$82 & &\\
SL\,674   &9.45$\pm$0.05 &3.6$\pm$0.2 & 6.9$\pm$0.3 & 29.1$\pm$2.4 & 22.7 & 15.7$\pm$10.9 & 536 & 204$\pm$70 & &\\
SL\,769   &9.35$\pm$0.10 &4.8$\pm$0.2 & 8.2$\pm$0.3 & 26.7$\pm$2.4 & 21.5 & 17.7$\pm$12.2 & 721 & 336$\pm$83 & &\\
LW\,477   &9.10$\pm$0.05 &2.7$\pm$0.5 & 6.3$\pm$0.6 & 26.7$\pm$2.4 & 21.6 &  2.2$\pm$1.5  & 155 & 308$\pm$25 & &\\
LW\,483   &9.10$\pm$0.05 &2.7$\pm$0.5 & 6.3$\pm$0.6 & 29.1$\pm$2.4 & 32.0 &  5.5$\pm$3.8  & 163 & 299$\pm$21 & &\\
\hline
\multicolumn{11}{c}{Additional clusters taken from the literature} \\
\hline
NGC\,411  & 9.16$\pm$0.02 & 4.23$\pm$0.26 & 6.12$\pm$0.79 & & & 4.67$\pm$0.03 & & & 516 & 2\\
NGC\,419  & 9.16$\pm$0.02 & 5.48$\pm$2.01 & 7.67$\pm$2.86 & & & 5.38$\pm$0.08 & & & 560 & 2\\
NGC\,1651 & 9.30$\pm$0.02 & 4.57$\pm$0.36 &12.82$\pm$2.01 & & & 4.91$\pm$0.06 & & & 315 & 2\\
NGC\,1718 & 9.25$\pm$0.02 & 3.74$\pm$0.24 & 5.42$\pm$0.56 & & & 4.83$\pm$0.07 & & & 406 & 2\\
NGC\,1751 & 9.15$\pm$0.02 & 5.76$\pm$0.41 & 7.10$\pm$0.87 & & & 4.81$\pm$0.06 & & & 353 & 2\\
NGC\,1755 & 7.9$\pm$0.05  &  1.7          &               & & &  4.0          & & &  25 & 6,7,8 \\
NGC\,1783 & 9.23$\pm$0.02 &10.50$\pm$0.49 &11.40$\pm$2.24 & & & 5.42$\pm$0.11 & & & 403 & 2\\
NGC\,1795 & 9.15$\pm$0.05 & 4.13$\pm$0.61 & 7.47$\pm$1.23 & & & 4.50$\pm$0.07 & & & 350 & 1\\ 
NGC\,1806 & 9.20$\pm$0.02 & 5.91$\pm$0.27 & 9.04$\pm$1.24 & & & 5.10$\pm$0.06 & & & 370 & 2\\
NGC\,1846 & 9.23$\pm$0.02 & 8.02$\pm$0.49 & 8.82$\pm$0.68 & & & 5.24$\pm$0.09 & & & 567 & 2\\
NGC\,1850 & 8.00$\pm$0.05 & 2.7           &   11.2        & & &  5.3          & & & 44.4 & 4,5\\ 
NGC\,1852 & 9.15$\pm$0.02 & 5.10$\pm$0.46 & 6.97$\pm$0.83 & & & 4.66$\pm$0.07 & & & 312 & 2\\
NGC\,1856 & 8.48$\pm$0.10 & 3.18$\pm$0.12 & 8.00$\pm$0.90 & & & 5.00$\pm$1.00 & & & 150 & 3\\
NGC\,1978 &  9.3          &   4.3          &     9.6        & & &    5.5         & & &  $<$100 & 2,9,10,11\\
NGC\,1987 & 9.04$\pm$0.02 & 4.18$\pm$0.46 &12.78$\pm$3.05 & & & 4.74$\pm$0.04 & & & 234 & 2\\
NGC\,2108 & 9.00$\pm$0.025& 5.42$\pm$0.27 & 7.20$\pm$0.76 & & & 4.71$\pm$0.07 & & & 230 & 2\\
NGC\,2154 & 9.19$\pm$0.02 & 4.50$\pm$0.29 & 5.69$\pm$0.51 & & & 4.61$\pm$0.06 & & & 431 & 2\\
LW\,431   & 9.28$\pm$0.02 & 4.03$\pm$0.24 & 9.10$\pm$3.16 & & & 4.56$\pm$0.07 & & & 277 & 2\\
Hodge\,2  & 9.35$\pm$0.02 & 2.67$\pm$0.41 & 9.09$\pm$2.33 & & & 4.70$\pm$0.07 & & & 363 & 2\\
IC\,2146  & 9.28$\pm$0.05 & 8.89$\pm$1.36 &12.53$\pm$1.92 & & & 4.49$\pm$0.07 & & & 410 & 1\\\hline

\end{tabular}

\noindent References: 1) \citet{cetal14}; 2) \citet{goudfrooijetal14}; 3) \citet{miloneetal15};
4) \citet{mvdm05}; 5) \citet{bastianetal16}; 6) \citet{elsonetal89}; 7) \citet{poetal12}; 
8) \citet{miloneetal16}; {9) \citet{fischeretal1992}; 10) \citet{mietal09}; 11) \citet{mucciarellietal2007}}.
\end{table*}

\subsection{Cluster eMSTOs}
The width of the cluster MSTOs were measured following the precepts outlined in 
\citet[e.g.][]{goudfrooijetal11}, i.e., by building a histogram of the number of cluster stars
--previously decontaminated from field stars-- located
within a strip running perpendicular to the cluster MSs at the TO positions 
\citep[see also][]{p13,pb16,bastianetal16}. Such a
direction is called the MSTO axis. The cluster MSTOs were cleaned from field stars by using the 
field star cleaning procedure introduced by \citet{pb12} and successfully applied in 
different star field crowdness conditions \citep[e.g.][and references therein]{p14,piattietal15b,p16}. 
{The left panel of Fig.~\ref{fig:fig4} illustrates these procedures, in which we show the
field star cleaned CMD of NGC\,2203 and a field star CMD with the same area represented by
black and green dots, respectively. The magenta parallelogram represents the MSTO strip.
When building the histograms we have taken into account the projection of the photometric errors 
of each star along the MSTO axis by using a density estimator algorithm descrided in, 
for instance, \citet{pg13};
thus reproducing the real spread caused by photometric uncertainties. We also built MSTO
histograms using the same density estimator 
algorithm without considering photometric errors. When the latter are compared to
those including the effects of photometric uncertainties we found similar distributions.
Fig.~\ref{fig:fig5} depicts both histograms for the cluster sample, which suggests that the 
observed MSTO width is not driven by photometric errors.
We recall that the data sets of the studied clusters were obtained from identical observational 
setups (same telescope, detector, filters) and show tight similitudes of 
completeness and photometric error distributions as LW\,477 and LW\,483 
\citep[][see their Fig. 5]{pb16}.

We then fitted Gaussian functions to the cluster MSTO histograms
using the IRAF\footnote{IRAF is distributed by the National 
Optical Astronomy Observatories, which is operated by the Association of 
Universities for Research in Astronomy, Inc., under contract with the National 
Science Foundation.} {\sc ngaussfit} task and  the results are 
overplotted on Fig.~\ref{fig:fig5} using black lines.

\newpage

\subsection{Synthetic cluster experiments}

In order to estimate the true spread across the MSTO we need to correct for the effects of photometric errors and stellar binarity.  To do so, we created artificial clusters using the same isochrones and distance moduli that were used in the observational analysis of the clusters in our sample.  We stochastically selected 10,000 stars from a \citet{salpeter55} initial mass function between 1~\msun and the maximum stellar mass still alive at a given age.  We then added photometric errors as well as the effects of binarity, where we assumed a flat binary distribution.  In order to estimate the photometric errors and binary fraction we compared the width and distribution of stars across the (verticalised) main sequence \citep{pb16}. The photometric error was estimated by fitting a Gaussian to the main peak of the distribution, while the binary fraction was estimated from the extended tail of the distribution to redder colours.  We found that the observed distributions were well fit with simulations that had a photometric error in magnitude of 0.035~mag in both the $g$ and $i$ filters.  Our experiments led to high estimates of the binary fraction ($25\%$ binaries with mass ratios above 0.67), which is about twice that found in \citet{mietal09}.  We note that by adopting potentially higher binary fractions than are actually present, we will overestimate the inferred age spread of the synthetic cluster (i.e., SSP) which will cause our estimated age spreads of the observations to be underestimated.  However, in most cases this will not lead to a large change in the estimate spreads.

\begin{figure*}
	\includegraphics[width=\textwidth]{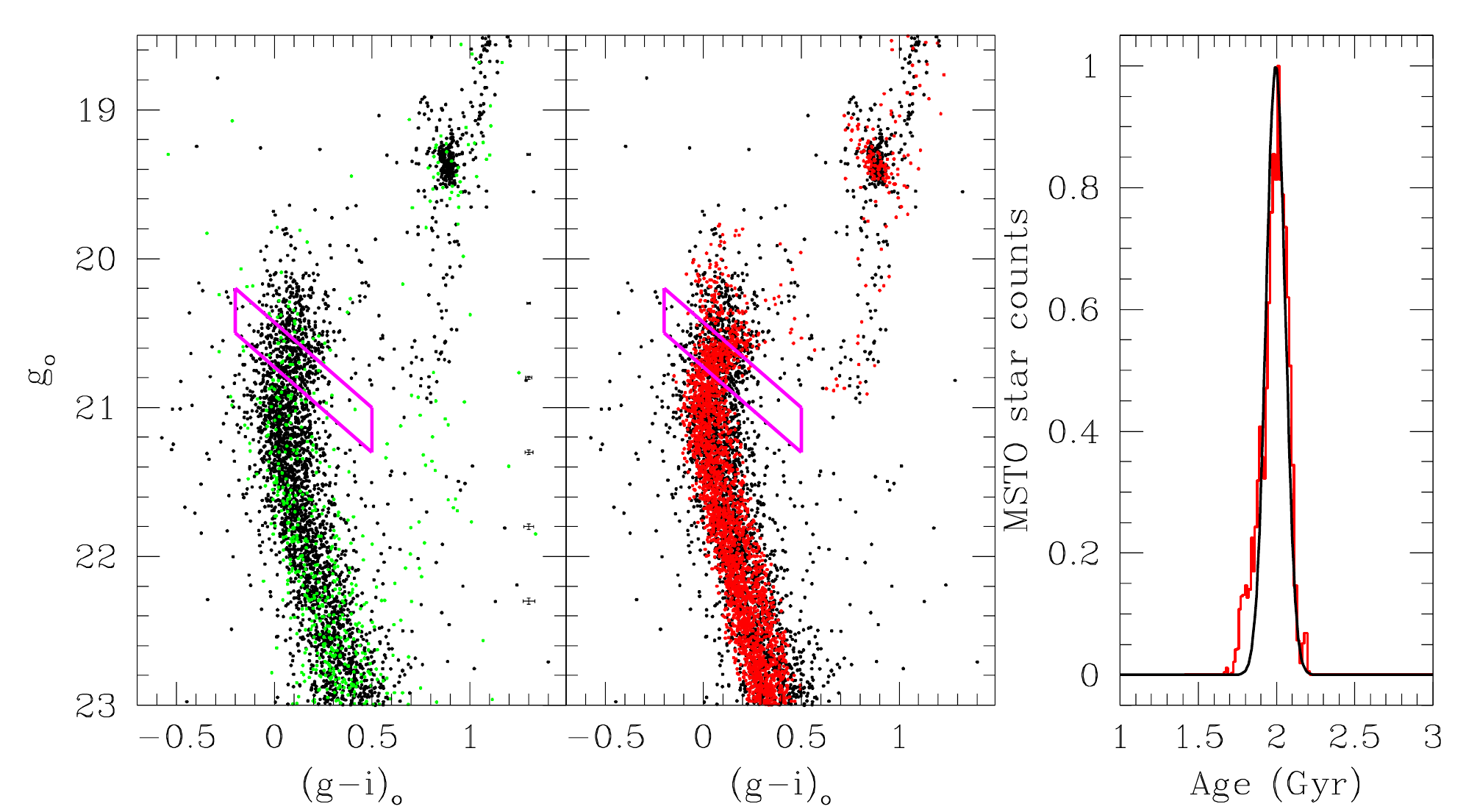}
    \caption{{\it Left panel:} Cleaned CMD of NGC\,2203 (black dots) and that for the field star 
CMD (i.e., the background with an equal area) (green dots). The MSTO strip is represented by a magenta
parallelogram. Errorbars are also drawn in the right margin. {\it Middle panel:} Cleaned and
modelled CMDs drawn with black and red dots, respectively. {\it Right panel:} The resulting
inferred age distribution of synthetic stars located within the MSTO strip (red line) and the
respective fitted Gaussian function superimposed (black line).}
   \label{fig:fig4}
\end{figure*}

Each synthetic cluster was then analysed in the same way as the observed clusters.  An example is shown in Fig.~\ref{fig:fig4}, where in the middle panel the synthetic cluster stars are shown in red and the resulting inferred age distribution is shown in red in the right panel. For each observed cluster, we estimated the inferred age spread by approximating the observed and modelled (as an SSP of the best fitting age) age distributions as Gaussians, and subtracted the modelled width ($\sigma_{\rm SSP}$) from the observed width ($\sigma_{\rm obs}$) in quadrature to obtain the intrinsic inferred width ($\sigma_{\rm intrinsic}$). Fig.~\ref{fig:fig6} shows the observed (cleaned from field stars),
modelled and true $FWHM$$_{\rm MSTO}$ as a function of the cluster age, depicted with blue open triangles, red open boxes and black filled circles, respectively.

We note that our estimated widths are in good agreement with that estimated in the literature (based mostly on HST observations), but there are some outliers.  In particular, our estimate of the width of NGC~2249 ($\sim170$~Myr) is significantly less than that found by 
\citet[][- $\sim450$~Myr]{cetal14}.  Part of this difference may be due to the fact that we assign a larger age (1.4~Gyr) to the cluster than Correnti et al. (1~Gyr).  However, we note that such a large inferred age spread is clearly incompatible with our observed CMD of this cluster.

As a test of the adopted method, we have estimated the inferred age spread of NGC~2203, based on the HST/WFC3 photometric catalogue of \citet{goudfrooijetal14}.  We find an age spread of 580~Myr (without correcting for the effects of photometric errors or binarity), in decent agreement with that found by \citet{goudfrooijetal14} and \citet{niederhoferetal16}, although significantly larger than we found based on our ground based data ($\sim350$~Myr).  The source of this discrepancy is unknown. We speculate in the possibility that the larger GMOS-S field of view could allow us
to do a better job in cleaning the cluster CMDs from field star contamination. On average, our estimated FWHM of the inferred age spread is $\sim50$~Myr less than that found in the literature.

\begin{figure*}
	\includegraphics[width=\textwidth]{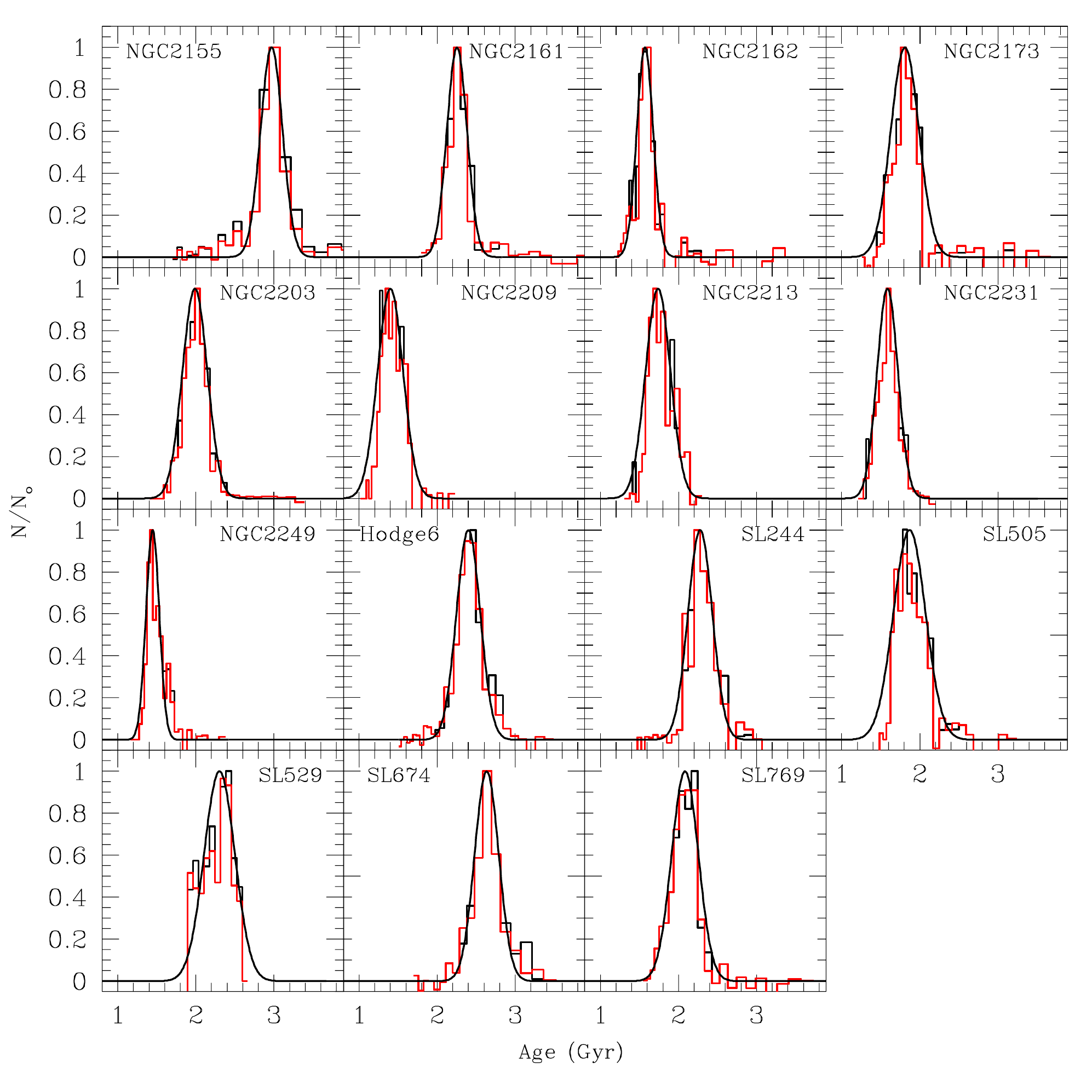}
    \caption{Normalised star counts along the MSTO axis \citep[see][]{goudfrooijetal11}. 
Thick black and red histograms refer to the star counts obtained by considering or not
the photometric uncertainties, respectively. Gaussian distributions fitted to the black histograms
are also superimposed.}
   \label{fig:fig5}
\end{figure*}

\begin{figure*}
	\includegraphics[width=\textwidth]{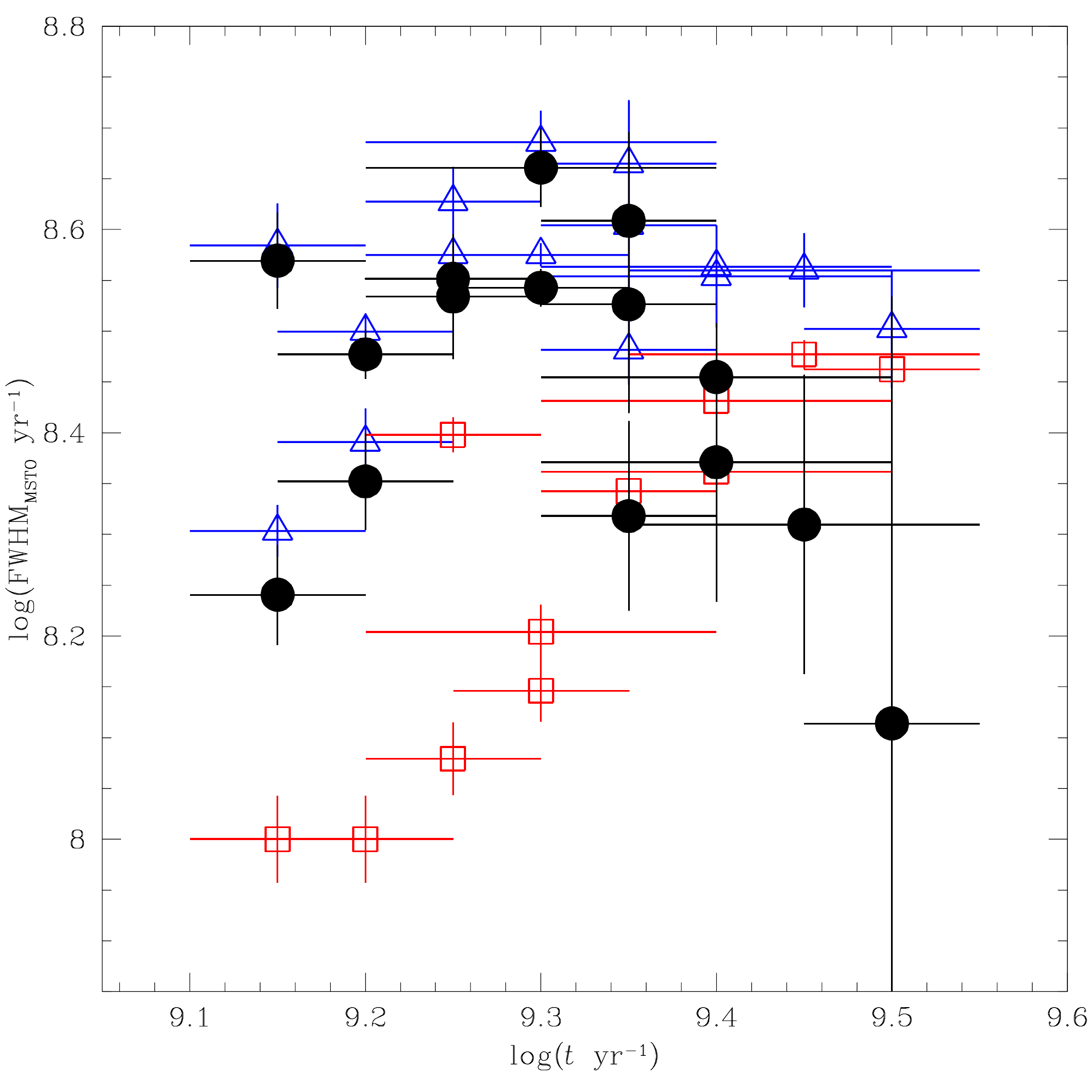}
    \caption{The observed (cleaned from field stars), modelled and true $FWHM$$_{\rm MSTO}$ 
as a function of the cluster age, depicted with blue open triangles, red open boxes and black 
filled circles, respectively. Errorbars are also drawn.}
   \label{fig:fig6}
\end{figure*}

 For comparison purposes, 
Table~\ref{tab:table1} includes in its last columns the $FWHM$s (eMSTOs) taken from \citet{cetal14} and 
\citet{goudfrooijetal14}, respectively.  As can be seen, there is
a general good agreement with our values, although some differences are expected to arise due to
different cluster regions and the use of ground based and space based images.
In the following, we use our derived values on a homogeneous basis.

\section{Analysis and Discussion}
\label{sec:discussion}
The cluster parameter values listed in Table~\ref{tab:table1} allow us to 
carry out an analysis in the light of some recent results about the origin and astrophysical
scenarios in which eMSTOs can arise. We enlarged our sample by adding those clusters
analysed by \citet{cetal14} and \citet{goudfrooijetal14}.  We used their ages, $r_c$,
$FWHM$$_{\rm MSTO}$ and masses. We also included NGC\,1755
\citep{elsonetal89,poetal12,miloneetal16}, NGC\,1850 \citep{mvdm05,bastianetal16} and 
NGC\,1856 \citep{cetal14,miloneetal15}, three young LMC clusters with confirmed eMSTOs. 
We did not include some clusters studied by \citet{mietal09} because of the lack of 
known structural parameters.
This is the largest cluster sample so far used to study the eMSTO phenomemon. As far
as we aware of, it consists of all the available literature data.

\citet[][hereafter K12]{kelleretal12} confirmed results by \citet{kelleretal11} in the sense that
clusters with eMSTO morphology have the largest core radii at a given age. 
According to the authors, the correspondence between the large cluster radius and the
presence of eMSTOs  implies that either the formation of stars subsequent to the primary stars 
of the cluster is fostered by the environment of a cluster of large core radius
or it imparts a kinematic signature that acts to increase the core radius of the host cluster. 
They drew such conclusions by linking results of \citet{mackeyetal08b} -- who
 showed that primordial mass segregation leads to a cluster
of $\sim$ 1-2 Gyr have a larger core radius than unsegregated ones -- and
their Fig. 4. Hence, they proposed that the eMSTO phenomenon is a common 
pathway for massive star clusters.

Fig.~\ref{fig:fig7} shows at the upper-left panel the region where K12 identified eMSTO clusters
depicted by a thick contour rectangle. Clusters taken from the literature 
\citep{cetal14,goudfrooijetal14} were drawn with solid black boxes to differentiate them from our
own cluster sample (large coloured boxes). As can be seen,
the marked rectangular area is mostly occupied by clusters previously analysed by K12; 
NGC\,2209 ($\log(t$ yr$^{-1}) =$ 9.15) being the only cluster studied here. 
NGC\,2209 complies with the K12's prediction in the sense that an eMSTO 
cluster does have a large core radius, since it lies at the top of that rectangular area and
observations show that it has an eMSTO \citep{cetal14}. Likewise, we here
identified other clusters (NGC\,2155, 2161, 2173, 2203, Hodge\,6, SL\,674 and 769) with
 large core radii exhibiting eMSTOs as shown in the upper-right panel.

\begin{figure*}
	\includegraphics[width=\textwidth]{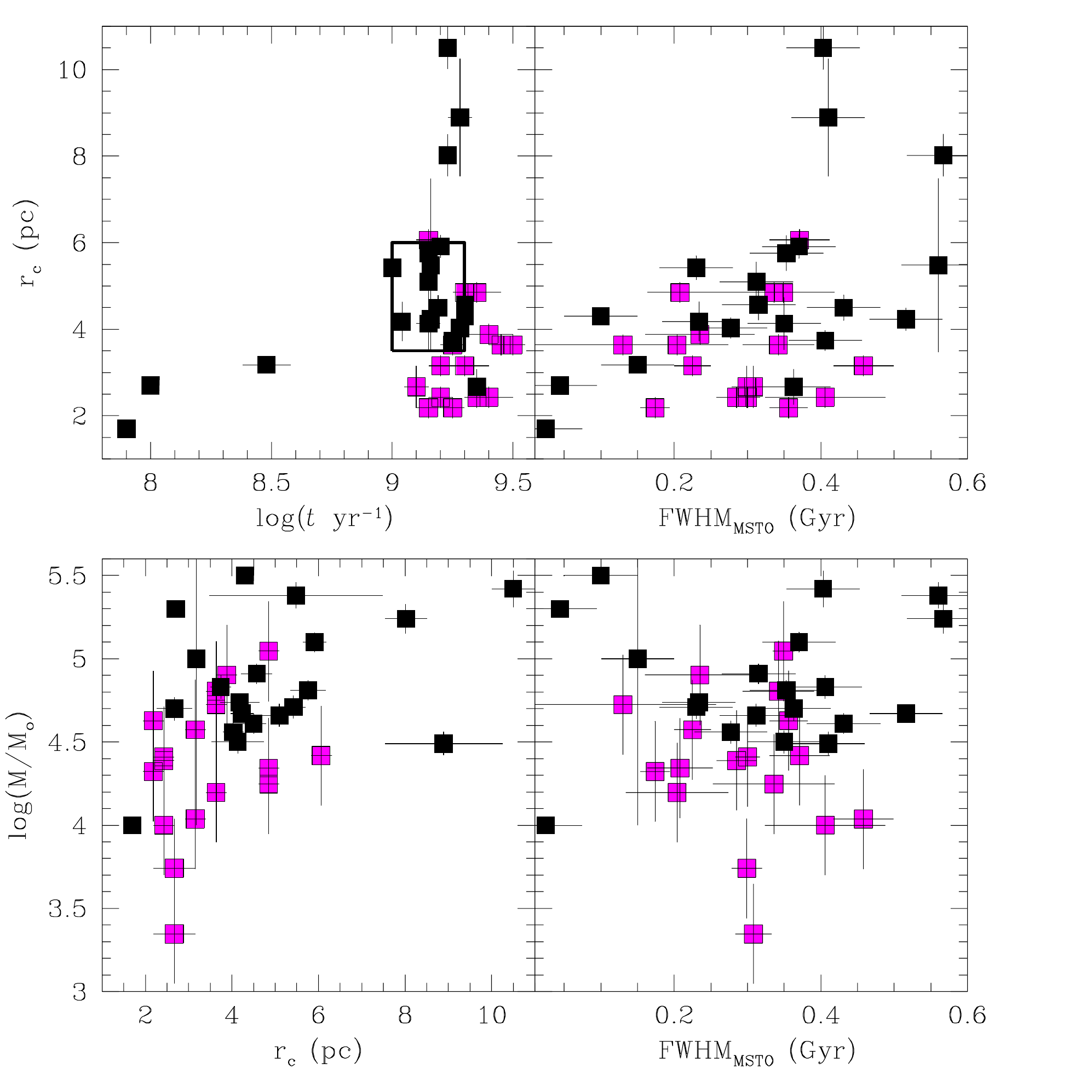}
    \caption{Relationships between various cluster parameters. 
Black solid boxes represent additional clusters taken from the 
literature (see Table~\ref{tab:table1}).
}
   \label{fig:fig7}
\end{figure*}

\begin{figure*}
	\includegraphics[width=\textwidth]{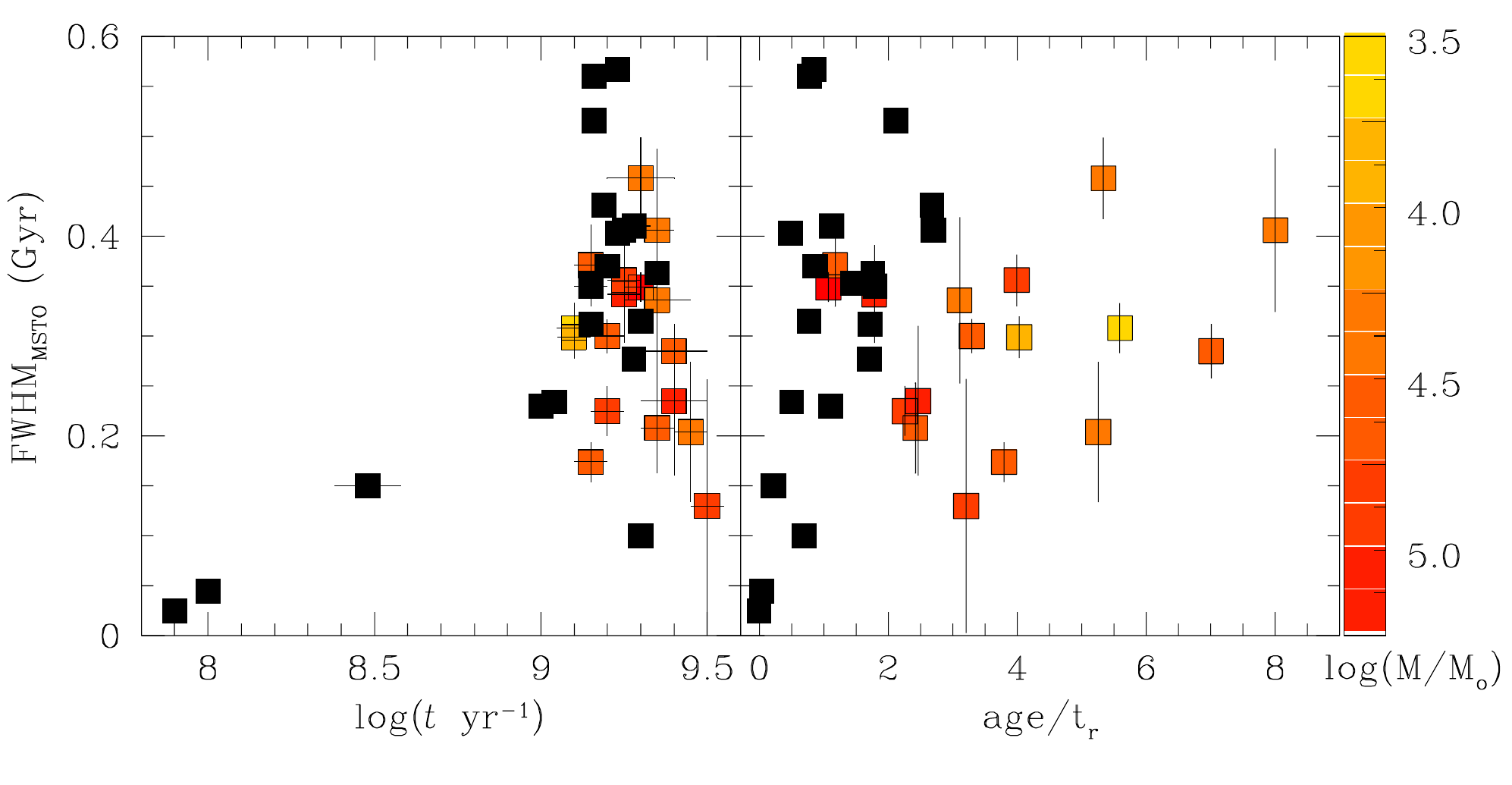}
    \caption{Relationships between various cluster parameters. 
Black solid boxes represent additional clusters taken from the 
literature (see Table~\ref{tab:table1}).
}
   \label{fig:fig8}
\end{figure*}

However, and perhaps more importantly, we also identify in the upper-right panel a number of
 eMSTO clusters with smaller core radii (NGC\,2162, 2213, 2231, 2249, SL\,244, 505,
529, LW\,477 and 483). Looking at the upper-left panel, it appears that the eMSTO clusters in 
our sample spread all along the core radius range at the considered age regime. 
Notice that K12 considered NGC\,2173, 2213 and 2249 as intermediate-age clusters without any
hint of eMSTOs, while \citet[][hereafter G14]{goudfrooijetal14} and \citet{cetal14} 
did confirm their
eMSTO nature in good agreement with the present values.   Hence, our sample does not confirm 
that large core radii are related to the eMSTO phenomenon.

K12 also claimed that only clusters 
of mass similar to, or greater than, those now exhibiting the eMSTO at an age of 1--3 Gyr
(log($M_{cls}$/$M_\odot$) $\ga$ 4.8) possessed a sufficiently deep potential to capture and 
hold enriched gas for the formation of second-generation stars, and therefore
to present eMSTOs. 
However, the bottom-panels of Fig.~\ref{fig:fig7} suggests
not only massive clusters can host eMSTOs.

G14 found  that the width of the MSTO region
correlates with the cluster's current mass with a probability of the absence of a correlation
smaller than 1.3 per cent. They noted, however, that due to different ages and radii clusters
underwent different amount of mass loss. Particularly, they applied "correction factors" to boost the initial escape velocity by adopting extreme mass loss and radii expansion. 
From Table~\ref{tab:table1}, however, we learn 
rather that only clusters with masses larger than $\log(M_{cls}/M\odot)$ $\sim$ 4.3 follows
the relation fitted by G14 (see their Fig. 5), and that less massive 
clusters -- most of them having relatively large $FWHM_{\rm MSTO}$ values -- are 
far from 
that straight line (see also bottom-right panel of Fig.~\ref{fig:fig7}). 
Since cluster masses 
and escape velocities are largely interchangeable, our
results thus contradict argumentations by G14, in the sense that eMSTOs only occur for clusters 
whose early escape velocities are higher than the wind velocities
of stars that provide material from which second-generation stars can form.

In Fig.~\ref{fig:fig8}, we show 
the $FWHM_{\rm MSTO}$ versus $\log(t$ yr$^{-1})$ for the whole cluster sample.
We included the cluster masses as a third parameter, distinguishing them
with a colour-coded scale as shown at the right-hand border of the figure. 
The relationship $FWHM_{\rm MSTO}$ versus $\log(t$ yr$^{-1})$ 
seems to follow a inverted-V trend, peaked at  $\log(t$ yr$^{-1})$ $\sim$ 9.2-9.3 (age $\sim$
2 Gyr), in very good agreement with \citet[][see their Fig. 4]{bastianetal16}.
For clusters younger than $<1$~Gyr there is a stronger correlation as found by
\citet{niederhoferetal15b} and discussed at length in \citet{niederhoferetal16}.
Notice that our 
Fig.~\ref{fig:fig8} (right-hand panel) reveals that eMSTO features are found
in clusters with different relaxation times.

Finally, of particular note is the high mass cluster NGC\,1978, which does not show
any evidence of an age spread \citep[$<100$~Myr][]{mietal09}.  It’s mass 
\citep[logm$/M_{\odot} \sim5.5$][]{goudfrooijetal14} and escape velocity suggest that it
should show a very large age spread in the G14 scenario, in contradiction to the
observed tight eMSTO.  Alternatively, since its age is past the `turn-over` in the
age-delta(age) plot, it would be expected to have a very narrow MSTO in the stellar
rotation scenario, consistent with what is observed.  Taken together with the
previous evidence, this suggests that actual age spreads are not present in massive
clusters.

\section{Conclusions}
\label{sec:conclusions}

From the photometric data sets of 15 LMC clusters observed by \citet{petal14}
with the Gemini-South telescope, we derived their structural parameters, masses,
relaxation times and $FWHM_{\rm MSTO}$ values from a fully consistent
method to that described by \citet{pb16} for other two low-mass LMC clusters,
also included in our subsequent analysis. The $FWHM_{\rm MSTO}$ values do not
take into acount the effect of stellar binarity and cover the range $\sim$ 
200--500 Myr, i.e., they do show eMSTOs. Six
clusters in common with \citet{cetal14}'s and G14's samples of eMSTO clusters show 
an overall good agreement of their core radii and $FWHM_{\rm MSTO}$ values, so that
we enlarged our cluster sample by adding those clusters studied by them. We gathered
in total 37 eMSTO LMC clusters. This is the largest cluster sample so far used to study
the eMSTO phenomenon.

The derived cluster core radii spread over the range of values ($\approx$ 4--6 pc)
claimed by K12 to exhibit eMSTOs. However, our whole core radii range reaches
values as small as $\sim$ 2 pc, thus providing the first evidence that they are 
not related to eMSTOs. Since large core radii and large masses are roughly
correlated, K12 also suggested a lower mass limit of log($M_{cls}$/$M_\odot$) $\sim$ 4.8
for eMSTO clusters. Here we found that clusters with masses $\ga$ 10$^{3.35}$ 
also host eMSTOs. Likewise, this result does not support the scenario suggested by G14 
about that eMSTOs only occur for clusters with high early escape velocities,  since 
cluster masses and escape velocities are interchangeable. Moreother, if eMSTOs
were orginated by two generations of stars, they should mainly retain second-generation
stars after evolving several relaxation times. However, we show that eMSTOs are
observed in dynamically relaxed clusters. 

Finally, we confirm from a larger cluster sample the type of ``$\Lambda$" distribution 
in $FWHM_{\rm MSTO}$ versus $\log(t$ yr$^{-1})$ plane found by \citet{niederhoferetal16}. 
The way to test if age is the defining parameter of eMSTOs is to look for them
at younger clusters ($<$500 Myr). Indeed, NGC\,1755 \citep[][$\sim80$~Myr]{miloneetal16},
NGC\,1850 \citep[][$\sim100$~Myr]{bastianetal16}, 
NGC\,1856 \citep[][$\sim300$~Myr]{cetal14} show small eMSTOs, consistent with the trend.

\section*{Acknowledgements}
We thank the anonymous referee whose comments and suggestions
allowed us to improve the manuscript.  We thank Paul Goudfrooij for providing his photometric catalogues for NGC~2203.  NB gratefully acknowledges financial support from the Royal Society (University 
Research Fellowship) and the European Research Council (ERC-CoG-646928, Multi-Pop).




\bibliographystyle{mnras}

\input{paper.bbl}








\bsp	
\label{lastpage}
\end{document}